\documentclass{ws-ijmpd}
\usepackage{psfig}
\begin{document}

\markboth{N F Naidu, M Govender and K S Govinder} {Thermal evolution
of a radiating anisotropic star with shear}

%
\catchline{}{}{}{}{}
%

\title{Thermal evolution of a radiating anisotropic star with shear}
\author{N F Naidu \footnote{E-mail: 203507365@ukzn.ac.za}}
\author{M Govender \footnote{E-mail: govenderm43@ukzn.ac.za}}
\author{K S Govinder \footnote{E-mail: govinder@ukzn.ac.za}}
\address{Astrophysics and Cosmology Research Unit, School of Mathematical Sciences,
University of KwaZulu-Natal, Durban 4041, South Africa}

\maketitle

\begin{history}
\received{Day Month Year}
\revised{Day Month Year}
\end{history}

\begin{abstract}
We study the effects of pressure anisotropy and heat dissipation in
a spherically symmetric radiating star undergoing gravitational
collapse. An exact solution of the Einstein field equations is
presented in which the model has a Friedmann-like limit when the
heat flux vanishes. The behaviour of the temperature profile
of the evolving star is investigated within the framework of causal
thermodynamics. In particular, we show that there are significant
differences between the relaxation time for the heat flux and the
relaxation time for the shear stress.
\end{abstract}


\section{Introduction}    

It has been shown that the role of anisotropy can alter the
evolution and subsequently the physical properties of stellar
objects. As an example, investigations have shown that the maximal
surface redshift for anisotropic stars may differ drastically from
isotropic stars. The origin of anisotropy within the stellar core
has received widespread attention amongst astrophysicists. A
review of the origins and effects of local anisotropy in stellar
objects was carried out by Herrera and Santos\cite{hers}, and Chan
{\em et al}\cite{chan1} and more recently by Herrera {\em et
al}\cite{hers1}. The physical processes that are responsible for
deviations from isotropy can be investigated in the high and low
density regimes. Hartle {\em et al}\cite{hart} have shown that
pion condensation at nuclear densities ($0.2 f^{-3} < \rho < 2.0
f^{-3}$, $1 f = 10^{-13}$ cm) can drastically reduce the pressure
and hence impact on the evolution of the collapsing star. At
higher densities the short range repulsion effects dominate which
damp out the pionic effects giving rise to significantly different
values for the pressure. As pointed out by Martinez\cite{mart1}
viscosity effects due to neutrino trapping at nuclear densities
can alter the gravitational collapse of a radiating, viscous star.
Anisotropic velocity distributions and rotation can induce local
anisotropy in low density systems.

The boundary of a radiating star divides spacetime into two
distinct regions, the interior and the exterior region. In order to fully describe the evolution of such a system, one needs to satisfy the junctions conditions for the smooth matching of the
interior and exterior spacetimes.  In the case of a shear-free, spherically
symmetric star undergoing dissipative gravitational collapse these conditions were
first derived by Santos\cite{santos}. Subsequently, many models of
shear-free radiative collapse were developed and the physical
viability of these models were studied in great
detail\cite{kolas,kram,gram2,gov1,gov2}. These junction conditions
were later generalised to include pressure anisotropy\cite{chan2}
and the electromagnetic field{\cite{ma}}.

There are a few exact solutions to the Einstein field equations for
a bounded shearing matter configuration. There have been numerous
attempts to produce models of radiative gravitational collapse which
incorporate the effects of shear\cite{barr,chan3,herr}. Most
treatments to date are based on numerical results as the resulting
temporal evolution equation derived from the junction conditions is
highly nonlinear\cite{chan3}. It is in this spirit that we seek an
exact solution of the Einstein field equations which represents a
spherically symmetric radiating star undergoing dissipative
gravitational collapse with nonzero shear.

In this treatment, we consider the relaxation effects due to the
heat flux and shear separately. We show that earlier assumptions of
the relaxation time being proportional to the corresponding
collision time only hold for a limited regime of the evolution of
the star. These results agree with earlier suggestions by Anile {\em
et al}\cite{anile}. We are also in a position to integrate the
causal heat transport equation and obtain the corresponding causal
temperature profile in the interior of the star.

This paper is organised as follows. In section 2, we provide the
Einstein field equations for the most general, nonrotating,
spherically symmetric line element. The energy momentum tensor for
the interior spacetime is that of an imperfect fluid with heat
conduction and pressure anisotropy. In section 3, the junction
conditions required for the smooth matching of the interior and
exterior spacetimes across a time-like hypersurface are presented.
The transport equations for the dissipative fluxes within the
framework of causal thermodynamics are presented in section 4. The
general solution for the special case of constant mean collision
time in both the causal and noncausal theories are presented. The
junction conditions are used to generate a reasonable model of
radiating anisotropic collapse in section 5. When the heat flux
vanishes, our model tends towards a Friedmann-like limit, similar to
the result obtained for the shear-free case studied by Kolassis {\em
et al}\cite{kolas}. A discussion of some relevant physical
parameters is presented in section 6.

\section{Spherically symmetric
Spacetimes}

The interior spacetime of a non-rotating spherically symmetric
collapsing star is most generally described by the line element (in
coordinates $(x^a) = (t, r, \theta, \phi)$)
\begin{equation} \label{a9}ds^2 = -A^2dt^2 + B^2dr^2 + Y^2(d\theta^2
+ \sin^2{\theta}d\phi^2)\,,
\end{equation} where $A$, $B$ and $Y$ are functions of the coordinates $t$
and $r$. The fluid four--velocity ${\bf u}$ is comoving and is
given by
\begin{equation} u^a = \frac{1}{A}\delta^{a}_0 \,.
\end{equation} The kinematical quantities for the line element
(\ref{a9}) are given by \begin{eqnarray} \label{a10} {\dot u}^a
&=&
\left(0, \frac{A^{\prime}}{AB^2}, 0, 0\right) \label{a10b}\\ \nonumber \\
\Theta &=& \frac{1}{A}\left(\frac{\dot B}{B} + 2 \frac{\dot
Y}{Y}\right) \label{a10c},
 \end{eqnarray}
 where ${\dot u}^a$ is the
 four--acceleration vector and $\Theta$ is the expansion scalar. The nonzero components of
 the shearing tensor are given by\cite{chan1}
\begin{eqnarray} \label{shear} {\sigma}_{11} &=&
\frac{2B^2}{3A}\left(\frac{{\dot B}}{B} - \frac{{\dot
Y}}{Y}\right)\label{sheara}\\ \nonumber \\ {\sigma}_{22} &=&
-\frac{2Y^2}{3A}\left(\frac{{\dot B}}{B} - \frac{{\dot
Y}}{Y}\right)\label{shearb}\\ \nonumber \\ {\sigma}_{33} &=&
\sigma_{22} \sin^2{\theta} .\end{eqnarray} The magnitude of the
shear scalar is given by
\begin{equation} {\sigma_{ab}}{\sigma^{ab}} = \sigma =
-\frac{1}{3A}\left(\frac{{\dot B}}{B} - \frac{{\dot
Y}}{Y}\right)\,.\end{equation}
 For a relativistic fluid, the
 kinematical quantities are important for studying the evolution of the
 system.
 The interior matter distribution is described by the following
 energy--momentum tensor
 \begin{equation}
 T_{ab} = (\rho + p)u_a u_b + p g_{ab} + {\pi}_{ab} + q_a u_b
           + q_b u_a \label{u6}\,,
           \end{equation}
           where $p$ is the isotropic pressure, $\rho$ is the density of the
           fluid, ${\bf \pi}_{ab}$ is the stress tensor
           and $q_a$ is
           the heat flux vector.
The stress tensor takes the form
\begin{equation}
{\pi}_{ab} = (p_R - p_T)(n_an_b -
\frac{1}{3}h_{ab})\,,\end{equation} where $p_R$ is the radial
pressure, $p_T$ is the tangential pressure, and ${\bf n}$ is a unit
radial vector given by
\begin{equation}
n^a = \frac{1}{B}\delta_1^a\,. \end{equation} The isotropic pressure
$p$ is related to the radial pressure and the tangential pressure
via
\begin{equation}
p = \frac{1}{3}[p_R + 2 p_T]\,.
\end{equation}
The coupled Einstein field equations for the interior matter
distribution become \begin{eqnarray} \label{a18} \rho   &=&
\frac{2}{A^2}\frac{{\dot B}}{B}\frac{\dot Y}{Y} + \frac{1}{Y^2} +
\frac{1}{A^2}\frac{{\dot Y}^2}{Y^2}\nonumber \\ \nonumber \\
&&- \frac{1}{B^2} \left( 2\frac{Y^{\prime\prime}}{Y} +
\frac{{Y^{\prime}}^2}{Y^2} -
2\frac{B^{\prime}}{B}\frac{Y^{\prime}}{Y} \right) \label{a18a} \\ \nonumber \\
p_R  &=& \frac{1}{A^2} \left(-2\frac{\ddot Y}{Y} - \frac{{\dot
Y}^2}{Y^2} +
2\frac{\dot A}{A}\frac{\dot Y}{Y} \right) \nonumber \\  \nonumber  \\
&& + \frac{1}{B^2} \left(\frac{{Y^{\prime}}^2}{Y^2} +
2\frac{A^{\prime}}{A}\frac{Y^{\prime}}{Y} \right) - \frac{1}{Y^2}
\label{a18b}  \\  \nonumber \\
p_T  &=& -\frac{1}{A^2}\left(\frac{\ddot B}{B} - \frac{\dot
A}{A}\frac{\dot B}{B} + \frac{\dot B}{B}\frac{\dot Y}{Y} -
\frac{\dot A}{A}\frac{\dot Y}{Y} + \frac{\ddot Y}{Y}\right)
 \nonumber \\ \nonumber \\
 && +  \frac{1}{B^2}\left(\frac{A^{\prime\prime}}{A} -
 \frac{A^{\prime}}{A}\frac{B^{\prime}}{B} + \frac{A^{\prime}}{A}\frac{Y^{\prime}}{Y} -
 \frac{B^{\prime}}{B}\frac{Y^{\prime}}{Y} + \frac{Y^{\prime\prime}}{Y}\right) \label{a18c}
 \\ \nonumber \\ q &=& -\frac{2}{AB^2} \left(-\frac{{\dot Y}^{\prime}}{Y} + \frac{\dot
 B}{B}\frac{Y^{\prime}}{Y} + \frac{A^{\prime}}{A}\frac{\dot Y}{Y} \right) . \label{a18d}
 \end{eqnarray} In the above we have defined \begin{equation}q = q^1\,, \end{equation} where $q^a = (0, q^1, 0, 0)$. The field
 equations (\ref{a18a})--(\ref{a18d}) describe the gravitational interaction of a shearing
 matter distribution with heat flux and anisotropic pressure.

\section{Spherical collapse with heat flow}

The problem of gravitational collapse within the context of general
relativity was first investigated by Oppenheimer and
Snyder\cite{snyder} in which the interior matter distribution was
taken to be dust with the exterior spacetime being Schwarzschild.
With the discovery of the Vaidya solution\cite{vaidya}, it became
possible to study radiative gravitational collapse where the
collapsing core radiated energy to the exterior spacetime\cite{santos,kram,herr,bon1}.
The Vaidya solution which describes
the exterior spacetime of a radiating star is given by
\begin{equation} \label{v1} ds^2 = - \left( 1 -
\frac{2m(v)}{{\sf{r}}}\right) dv^2 - 2dvd{\sf{r}} + {\sf{r}}^2
\left(d\theta^2 + \sin^2\theta d\phi^2 \right)\,\end{equation} in
the coordinates $(x^a) = (v, {\sf r}, \theta, \phi)$. The quantity
$m$($v$) represents the Newtonian mass of the gravitating body as
measured by an observer at infinity. The metric (\ref{v1}) is the
unique spherically symmetric solution of the Einstein field
equations for radiation in the form of a null fluid. The Einstein
tensor for the line element (\ref{v1}) is given by
\begin{equation}  \label{v2} G_{ab} =
-\frac{2}{{\sf{r}}^2}\frac{dm}{dv} \delta^0_a \delta^0_b\,.
\end{equation} The energy--momentum tensor for null radiation assumes
the form
\begin{equation} \label{v3} T_{ab} = {\epsilon}w_{a}w_{b}\,,
\end{equation}
where the null four--vector ${\bf w}$ is given by $w_a = (1, 0, 0, 0)$. Thus
from (\ref{v2}) and (\ref{v3}) we have that \begin{equation}
\label{v4} {\epsilon} = -\frac{2}{{\sf{r}}^2}\frac{dm}{dv}
\end{equation} for the energy density of the null radiation. Since the
star is radiating energy to the exterior spacetime we must have
$\displaystyle\frac{dm}{dv} \leq 0 $.

In deriving the junction conditions we employ the Darmois conditions
as these were shown to be the most convenient and
reliable\cite{bon2}. Here we merely state the results of the
matching of the line elements (\ref{a9}) and (\ref{v1}) since these
conditions have been extensively investigated in the past. For a
more comprehensive treatment of the junction conditions the reader
is referred to the works of Glass\cite{glass2},
Chan\cite{chan2,chan3} and Govender {\em et al}\cite{gov2}.

We require that the metrics (\ref{a9}) and (\ref{v1}) match smoothly
across the boundary $\Sigma$. This generates the first junction
condition
\begin{equation}   \label{25} (ds^2_-)_{\Sigma} = (ds^2_+)_{\Sigma}
= ds^2_{\Sigma} \,.\end{equation} (We use the notation
$(\hspace{0.2cm})_{\Sigma} $ to represent the value of
$(\hspace{0.2cm})$ on $\Sigma$.) The second junction condition is
obtained by requiring continuity of the extrinsic curvature
across the boundary. This gives \begin{equation} \label{26}
K^+_{ij} = K^-_{ij}  \,,\end{equation} where
\begin{equation} \label{27} K^{\pm}_{ij} \equiv -n^{\pm}_a
\displaystyle\frac{{{\partial}^2\cal{X}}^{a}_{\pm}}{{\partial\xi}^{i}{\partial\
xi}^j} -n^{\pm}_{a} \Gamma^{a}{}_{cd}
\displaystyle\frac{{\partial\cal{X}}^{c}_{\pm}}{{\partial\xi}^{i}}\displaystyle
\frac{{\partial\cal{X}}^{d}_{\pm}}{{\partial\xi}^{j}}\,,
\end{equation} and $ n_{a}^{\pm}$(${\cal{X}}^b_{\pm}$) are the
components of the vector normal to $\Sigma$. We find that the
necessary and sufficient conditions on the spacetimes for the
first junction condition to be valid are that
\begin{eqnarray} \label{b15} A(r_{\Sigma}, t)dt &=&
\left( 1 - \displaystyle\frac{2m}{{\sf r}_{\Sigma}} +
2\displaystyle\frac{d{\sf r}_{\Sigma}}{dv} \right)^{\frac{1}{2}}dv
\label{b15a} \\ \nonumber \\ Y(r_{\Sigma}, t) &=& {\sf
r}_{\Sigma}(v) \label{b15b} .\end{eqnarray} By equating the
appropriate ext\-rinsic curvature components for the interior and
exterior spacetimes we generate the second set of junction
conditions which are given by \begin{eqnarray} \label{b20} m(v)
&=&
 \left[\frac{Y}{2}\left(1 + \frac{{Y_t}^2}{A^2} - \frac{{Y_r}^2}{B^2}
 \right)\right]_{\Sigma} \label{b20a} \\ \nonumber \\ (p_R)_{\Sigma} &=&
 (qB)_{\Sigma} \label{b20b} ,\end{eqnarray} where we may interpret $m(v)$ as representing the total
 gravitational mass within the surface $\Sigma$. The expression
 (\ref{b20a})
 corresponds to the mass function of Cahill and McVittie\cite{cah}
 representing
 spheres of radius $r$ inside $\Sigma$.

The important result $(p_R)_{\Sigma} = (qB)_{\Sigma}$, relating
the radial pressure $p_R$ to the heat flow $q$, was first
established by Santos\cite{santos} for shear--free spacetimes. The
first attempt to generalise the above junction conditions to
include shear for neutral matter was carried out by
Glass\cite{glass2}.

The equations (\ref{b15})--(\ref{b20b}) are the most general
matching conditions for the spherically symmetric spacetimes
${\cal{M}}^{+}$ and ${\cal{M}}^{-}$. Relation (\ref{b20b}) implies
that the radial pressure $p_R$ is proportional to the magnitude of
the heat flow $q$ which is non-vanishing in general.

The total luminosity for an observer at rest at infinity is given by
\begin{equation} \label{b25} L_{\infty}(v)  =
-\left(\displaystyle\frac{dm}{dv}\right)_{\Sigma} =
\frac{(p_R)_{\Sigma}}{2}\left[Y^2\left(\frac{Y^{\prime}}{B} + {\dot
Y}\right)^2\right]_{\Sigma} \,,\end{equation} where $\displaystyle\frac{dm}{dv}
\leq 0$ since $L_{\infty}$ is positive. An observer with
four--velocity $v^a = (\dot{v}, \dot{\sf{r}}, 0, 0)$ located on
$\Sigma$ has proper time $ \eta $ related to the time $ t $ by $ d
\eta = Adt $. The radiation energy density that this observer
measures on $ \Sigma $ is \begin{equation}{\epsilon}_{\Sigma} =
\frac{1}{4\pi}\left(-\displaystyle\frac{{\dot{v}}^2}{{\sf {r}}^2}
\displaystyle\frac{dm}{dv} \right)_{\Sigma} \,,\end{equation} and
the luminosity observed on $ \Sigma$ can be written as
\begin{equation}L_{\Sigma} = 4{\pi}{\sf{r}}^2{\epsilon}_{\Sigma} \,.
\end{equation} The boundary redshift $z_{\Sigma}$ of the radiation
emitted by the star is given by \begin{equation}  \label{b26} 1 +
z_{\Sigma} = \displaystyle\frac{dv}{d \eta}=
\left(\frac{Y^{\prime}}{B} + {\dot Y}\right)^{-1}_{\Sigma}
\end{equation} which can be used to determine the time of
formation of the horizon. The above expressions allow us to write
\begin{equation}1 + z_{\Sigma} =
\left(\frac{L_{\Sigma}}{L_{\infty}}\right)^{\frac{1}{2}}
\end{equation} which relates the luminosities $L_{\Sigma}$ to
$L_{\infty}$. The redshift for an observer at infinity diverges at
the time of formation of the horizon which is determined from
\begin{equation} \frac{Y^{\prime}}{B} + {\dot Y} =
0\,.\end{equation}

\section{Thermodynamics}
Previous treatments of heat flow in relativistic stellar models have
shown that relaxational effects play a significant role in the
evolution of the temperature and luminosity profiles during late
stages of collapse\cite{mart1,gov1,gov2,di}. Since we are interested
in the effects of shear within the stellar core, we employ the
causal transport equations for the heat flux and the shear stress.
Assuming there is no viscous/heat coupling, we have the following
relevant transport equations for the dissipative fluxes\cite{maart2}
\begin{eqnarray}
{\tau} h_a{}^b {\dot q}_{b} + q_a &=& -\kappa(D_a T + T {\dot
u}^a) - \left[\frac{1}{2}\kappa T^2\left(\frac{{\tau}_1}{\kappa
T^2}u^b\right)_{;b} q_a\right]\,\label{ctc}\\ \nonumber \\ {\tau}_1
h_a{}^ch_b{}^d {\dot {\pi}}_{cd} + {\pi}_{ab} &=&
-2\eta{\sigma}_{ab} - \left[\eta T\left(\frac{{\tau}_2}{2\eta
T}u^d\right)_{;d} {\pi}_{ab}\right]\,, \label{ctb} \end{eqnarray}
where ${\tau}$ and ${\tau}_1$ are the relaxation times of thermal
and shear viscous signals respectively. The truncated transport
equations, together with the no--coupling assumption are
differential equations of Maxwell--Catteneo form
\begin{eqnarray}
{\tau} h_a{}^b {\dot q}_{b} + q_a &=& -\kappa(D_a T + T {\dot
u}^a) \label{cmc}\\ \nonumber \\ {\tau}_1 h_a{}^ch_b{}^d {\dot
{\pi}}_{cd} + {\pi}_{ab} &=& -2\eta{\sigma}_{ab} \label{cmb}.
\end{eqnarray} Setting $\tau=\tau_1 = 0$ in the above, we regain the so-called Eckart
transport equations which predict infinite propagation velocities
for the dissipative fluxes.

In this paper we are primarily interested in heat transport in
relativistic astrophysics and hence (\ref{cmc}) plays a significant
role in determining the evolution of the causal temperature profile
of our models. For the line element (\ref{a9}) the causal transport
equation (\ref{cmc}) becomes \begin{equation} \label{ca1}
\tau(qB)_{,t} + A(qB) = -\kappa \frac{(AT)_{,r}}{B} \end{equation}
which governs the behaviour of the temperature. Setting $\tau = 0$
in (\ref{ca1}) we obtain the familiar Fourier heat transport
equation \begin{equation} \label{ca2} A(qB) = -\kappa
\frac{(AT)_{,r}}{B} \end{equation} which predicts reasonable
temperatures when the fluid is close to quasi--stationary
equilibrium.

For a physically reasonable model, we use the thermodynamic
coefficients for radiative transfer outlined in
Mart{\'i}nez\cite{mart1}. We consider the situation where energy is
carried away from the stellar core by massless particles that are
thermally generated with energies of the order of $kT$. The thermal
conductivity takes the form \begin{equation} \kappa =\gamma
T^3{\tau}_{\rm c} \label{a28}\,,\end{equation} where $\gamma$
($\geq0$) is a constant and ${\tau}_{\rm c}$ is the mean collision
time between the massless and massive particles. Based on this
treatment we assume the power--law behaviour
\begin{equation} \label{a29} \tau_{\rm c}
=\left({\alpha\over\gamma}\right) T^{-\omega} \,,\end{equation} where
$\alpha$ ($\geq 0$) and $\omega$ ($\geq 0$) are constants. With
$\omega={3\over2}$ we regain the case of thermally generated
neutrinos in neutron stars. The mean collision time decreases with
growing temperature, as expected, except for the special case
$\omega=0$, when it is constant. This special case can only give a
reasonable model for a limited range of temperature. Following
Mart{\'i}nez\cite{mart1}, we assume that the velocity of thermal
dissipative signals is comparable to the adiabatic sound speed which
is satisfied if the relaxation time is proportional to the collision
time: \begin{equation} \tau =\left({\beta \gamma \over
\alpha}\right) \tau_{\rm c} \label{a30}\,,\end{equation} where $\tau$
($\geq 0$)is a constant. We can think of $\tau$ as the `causality'
index, measuring the strength of relaxational effects, with $\tau=0$
giving the noncausal case.

Using the above definitions for $\tau$ and $\kappa$, (\ref{ca1})
takes the form \begin{equation} \beta (qB)_{,t} T^{-\omega} + A (q
B) = - \alpha \frac{T^{3-\omega} (AT)_{,r}}{B} \label{temp1}
\,.\end{equation} When $\beta=0$, we can find all noncausal
solutions of (\ref{temp1}), {\it viz.} \begin{eqnarray}
(AT)^{4-\omega}&=&\frac{\omega-4}{\alpha} \int A^{4-\omega} q B^2
{\rm d} r
+ F(t) \qquad \omega \ne 4\nonumber \label{noncausg}\\
\ln (AT) &=& - \frac{1}{\alpha}\int qB^2{\rm d} r + F(t)
 \qquad \omega=4 \label{noncaus4}
 ,\end{eqnarray}
 where $F(t)$ is an arbitrary function of integration.  This is fixed by
 the expression for the temperature of the star at its surface $\Sigma$.

In the case of constant mean collision time, {\it ie.} $\omega=0$,
the causal transport equation (\ref{temp1}) is simply integrated to
yield
\begin{equation} (AT)^4 = - \frac{4}{\alpha} \left[\beta\int A^3 B
(qB)_{,t}{\rm d} r + \int A^4 q B^2 {\rm d} r\right] + F(t)
\label{caus0} \end{equation} while one solution valid for a less
limited range of temperature can be found for $\omega=4$, which
corresponds to nonconstant collision time\cite{gov3}:
\begin{eqnarray} (AT)^4 &=& -\frac{4
\beta}{\alpha}\exp\left(-\int\frac{4qB^2}{\alpha} {\rm d} r\right)
\int A^3 B (qB)_{,t} \exp\left(\int\frac{4qB^2}{\alpha} {\rm d}
r\right){\rm d} r \nonumber \\
&&\mbox{}+ F(t) \exp\left(-\int\frac{4qB^2}{\alpha} {\rm d}
r\right) \label{caus4} .\end{eqnarray} In order to investigate the
relaxational effects due to shear we utilise (\ref{cmb}) as a
definition for the relaxation time for the shear stress. For the
metric (\ref{a9}) the shear transport equation (\ref{cmb}) reduces
to
\begin{equation}
{\tau}_1 = \frac{-P}{\dot{P} + \frac{8}{15}r_0\sigma
T^4},\end{equation} where we have used the coefficient of shear
viscosity for a radiative fluid\cite{maart2}
\begin{equation}\label{eta}
\eta = \frac{4}{15}r_0T^4\tau_1,\end{equation} $P=
\frac{1}{3}\left(p_T - p_R\right)$ and $r_0$ is the
radiation constant for photons. We have further assumed that
$\tau_1 = \beta_1 \tau_c$.
\section{Radiating anisotropic collapse}

 The junction condition $(p_R)_{\Sigma} = (qB)_{\Sigma}$ yields
\begin{equation}\left[ 2Y{\ddot Y} + {\dot Y}^2 - \frac{{Y^{\prime}}^2}{B^2} +
\frac{2}{B}Y{\dot Y}^{\prime} - 2\frac{{\dot B}}{B^2} Y Y^{\prime} +
1\right]_{\Sigma} = 0 \label{x},\end{equation} where we have set
$A(r,t) = 1$. A particular solution of (\ref{x}) is given by
\begin{eqnarray} \label{fun}
Y &=& rt^{2/3} \label{fun1}\\ \nonumber \\
B &=& \left( \frac{1 + c_1(r)e^{\frac{3t^{1/3}}{r}}}{1 -
c_1(r)e^{\frac{3t^{1/3}}{r}}}\right)t^{2/3} \label{fun2}
\end{eqnarray} which yields the line element \begin{equation}
ds^2 = -dt^2 + t^{4/3}\left[\left( \frac{1 +
c_1(r)e^{\frac{3t^{1/3}}{r}}}{1 -
c_1(r)e^{\frac{3t^{1/3}}{r}}}\right)^2dr^2 + r^2
d{\Omega}^2\right]\label{line}\,.\end{equation} With this form of
the line element, the field equations (\ref{a18})--(\ref{a18d})
become
\begin{eqnarray} \label{efe}\rho &=& \frac{4}{3\,t^2}\,\left( 1 +
\frac{1}{r^3} \left( \frac{r^2\,t^{\frac{1}{3}}}
          {1 - e^{\frac{3\,t^{\frac{1}{3}}}{r}}\,c_1(r)} +
         \frac{18\,t}
          {{\left( 1 +
               e^{\frac{3\,t^{\frac{1}{3}}}{r}}\,c_1(r)
               \right) }^3}
               \right. \right.\nonumber \\
&&\mbox{} \left.-
         \frac{3\,\left( r\,t^{\frac{2}{3}} + 9\,t
              \right) }{{\left( 1 +
               e^{\frac{3\,t^{\frac{1}{3}}}{r}}\,c_1(r)
               \right) }^2} +
         \frac{-\left( r^2\,t^{\frac{1}{3}} \right)  +
            3\,r\,t^{\frac{2}{3}} + 9\,t}{1 +
            e^{\frac{3\,t^{\frac{1}{3}}}{r}}\,c_1(r)}  \right)     \nonumber \\
&&\mbox{} \left.-
            \frac{3\,e^{\frac{3\,t^{\frac{1}{3}}}{r}}\,
         t^{\frac{2}{3}}\,
         \left( -1 + e^{\frac{3\,t^{\frac{1}{3}}}{r}}\,
            c_1(r) \right) \,c_1'(r)}{r\,
         {\left( 1 + e^{\frac{3\,t^{\frac{1}{3}}}{r}}\,
              c_1(r) \right) }^3} \right) \label{efe1}\\
p_R &=& \frac{-4\,e^{\frac{3\,t^{\frac{1}{3}}}{r}}\,c_1(r)}
  {t^{\frac{4}{3}}\,{\left( r +
        e^{\frac{3\,t^{\frac{1}{3}}}{r}}\,r\,c_1(r) \right)
        }^2} \label{efe2}\\
p_T &=& \frac{2}{3\,r^3\,t^{\frac{5}{3}}} \left(
\frac{-3\,r\,t^{\frac{1}{3}}}
       {{\left( -1 + e^{\frac{3\,t^{\frac{1}{3}}}{r}}\,
             c_1(r) \right) }^2} +
      \frac{r\,\left( 2\,r - 3\,t^{\frac{1}{3}} \right) }
       {-1 + e^{\frac{3\,t^{\frac{1}{3}}}{r}}\,c_1(r)} \right. \nonumber \\
&&\mbox{}+
      \frac{9\,t^{\frac{1}{3}}\,
         \left( 3\,t^{\frac{1}{3}}\,c_1(r) -
           r^2\,c_1'(r) \right) }{c_1(r)\,
         {\left( 1 + e^{\frac{3\,t^{\frac{1}{3}}}{r}}\,
              c_1(r) \right) }^2} +
      \frac{6\,t^{\frac{1}{3}}\,
         \left( -3\,t^{\frac{1}{3}}\,c_1(r) +
           r^2\,c_1'(r) \right) }{c_1(r)\,
         {\left( 1 + e^{\frac{3\,t^{\frac{1}{3}}}{r}}\,
              c_1(r) \right) }^3}  \nonumber \\
&&\mbox{}\left.+
      \frac{2\,r^2\,c_1(r) - 9\,t^{\frac{2}{3}}\,c_1(r) +
         3\,r^2\,t^{\frac{1}{3}}\,c_1'(r)}{c_1(r) +
         e^{\frac{3\,t^{\frac{1}{3}}}{r}}\,{c_1(r)}^2}
      \right) \label{efe3}\\
q &=& \frac{4\,e^{\frac{3\,t^{\frac{1}{3}}}{r}}\,c_1(r)\,
    \left( -1 + e^{\frac{3\,t^{\frac{1}{3}}}{r}}\,
       c_1(r) \right) }{r^2\,t^2\,
    {\left( 1 + e^{\frac{3\,t^{\frac{1}{3}}}{r}}\,
         c_1(r) \right) }^3}\label{efe4}.
\end{eqnarray}

In the absence of heat flux ($c_1 = 0$) our model yields
\begin{eqnarray}
Y &=& rt^{2/3}  \label{f1}\\ \nonumber \\
B &=& t^{2/3} \label{f2} \\ \nonumber \\
\rho &=& \frac{4}{3t^2} \label{f3} \\ \nonumber \\
p_R &=& p_T = 0 .\end{eqnarray} 
(Note that the solution presented above is not simply obtained by substituting $c_1 = 0$ into (\ref{efe1})--(\ref{efe4}).  Rather, one must go back to the original metric and then derive the field equations {\it ab initio}.) The above solution represents a dust
sphere and the metric is described by the Einstein--de Sitter
solution. We note that when $q = 0$ the pressure must vanish which
allows for the matter to have free-fall motion. For $q \neq 0$, the
pressure is non-vanishing so that it compensates for the outgoing
heat flux thus allowing for free-fall motion. With this in mind we
expect that the luminosity radius of the star in both the radiative
and non-radiative cases have the same temporal dependence.
Calculating the luminosity radius for our radiating model, we obtain
\begin{equation}Y_{\Sigma} = bt^{2/3}\end{equation} which is independent of $c_1$. Hence the case
$c_1 = 0$ reduces to the model investigated by Oppenheimer and
Snyder\cite{snyder}.

\section{Physical considerations}

In order to check the physical viability of our model, we
investigate the evolution of the temperature profile within the
framework of extended irreversible thermodynamics.

Utilising (\ref{caus0}) and (\ref{line}) we obtain
\begin{eqnarray} \label{tt1} T^4 &=& \frac{L_{\infty}}{(4\pi
\delta Y^2)_{\Sigma}} + \frac{16\beta c_1}{3\alpha
t^{5/3}}\left[\frac{e^{3t^{1/3}/b}}{b(1 + c_1e^{3t^{1/3}/b})} -
\frac{e^{3t^{1/3}/r}}{r(1 + c_1e^{3t^{1/3}/r})}\right] \nonumber \\ \nonumber \\
&& + \frac{16}{9\alpha t^2}\left\{\log{\left[\left(\frac{-1 +
c_1e^{3t^{1/3}/b}}{-1 + c_1e^{3t^{1/3}/r}}\right)^2 \left(\frac{1
+ c_1e^{3t^{1/3}/r}}{1 +
c_1e^{3t^{1/3}/b}}\right)^3\right]}\right\}\nonumber \\ \nonumber
\\ && +\frac{16}{3\alpha t}\left[\tanh^{-1}({c_1e^{2t^{1/3}/b}}) -
\tanh^{-1}({c_1e^{2t^{1/3}/r}})\right] .\end{eqnarray} where
$L_{\infty}$ is given by (\ref{b25}). We note that the causal and
the noncausal ($\beta = 0$) temperatures coincide at the boundary
($r = b$):
\begin{equation} T(t,b)=\tilde{T}(t,b) \,
\label{t1}\,.\end{equation}
\begin{figure}[t]
\centerline{ \psfig{figure=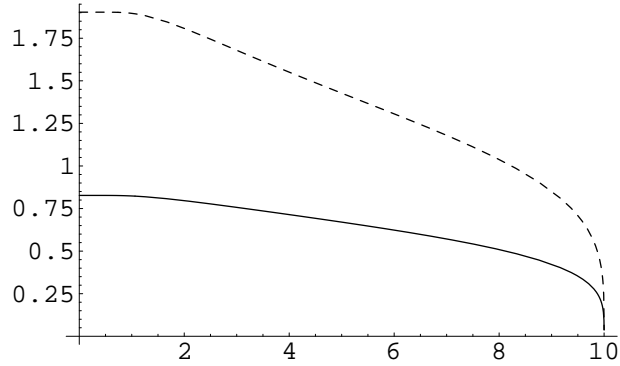}} \caption{Causal temperature
(dashed line), noncausal temperature (solid line) versus $r$.
\label{fig1}}
\end{figure}
\begin{figure}[t]
\centerline{ \psfig{figure=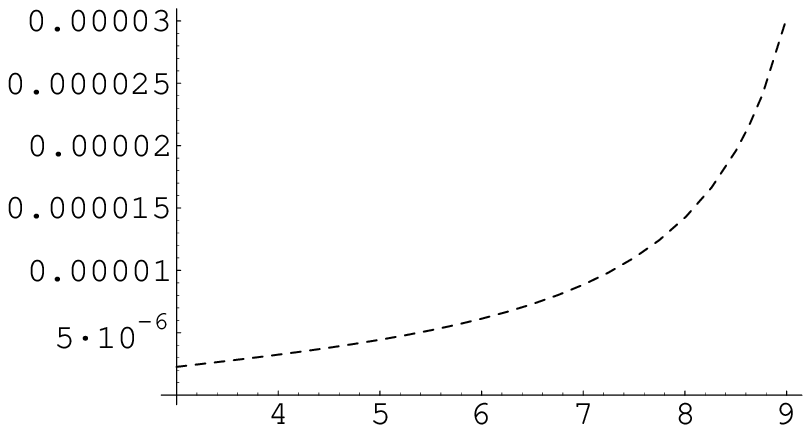} \psfig{figure=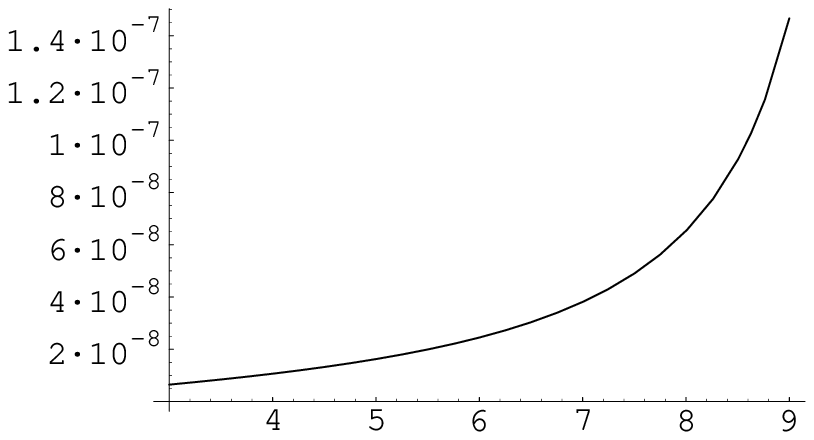}}
\caption{Relaxation time for the shear stress (close to equilibrium
- dashed line), (far from equilibrium - solid line) versus $r$.
\label{fig2}}
\end{figure}
 However, Figure \ref{fig1}  shows that at all interior points, the causal and
non-causal temperatures differ. In particular, we observe that the
causal temperature is greater than the non-causal temperature at
each interior point of the star. For small values of $\beta$, the
temperature profile is similar to that of the non-causal theory;
but as $\beta$ is increased, i.e. as relaxational effects grow, it
is clear from Figure \ref{fig1} that the temperature profile can
deviate substantially from that of the non-causal theory. Similar
results were obtained in the shear-free models studied by Herrera
and Santos\cite{hers} and Govender {\em et al}\cite{gov1,gov2}.
Also, from the plots in Figure \ref{fig2} the relaxation time for
the shear stress exhibits substantially different behaviour when
the fluid is close to hydrostatic equilibrium as opposed to
late-time collapse. In particular, we find that \begin
{equation}\frac{(\tau_{1})_{early}} {(\tau_{1})_{late}} \approx
100,\end{equation} emphasizing the importance of relaxational
effects during the different stages of collapse. We further note
that while the relaxation time for the heat flux is taken to be
constant, the relaxation time for the shear stress increases as
the collapse proceeds.

\begin{figure}[t]
\centerline{ \psfig{figure=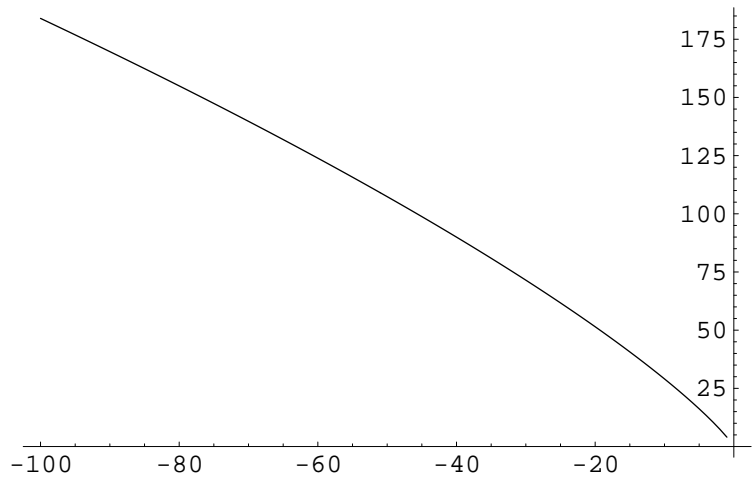}} \caption{Proper radius versus
time. \label{fig3}}
\end{figure}
Making use of (\ref{a9}) and (\ref{fun2}) the proper radius can be
written as
\begin{equation} \label{nolene74}
R = \int_0^{r_{\Sigma}}Bdr = t^{2/3}\int_0^{r_{\Sigma}} \left(
\frac{1 + c_1e^{\frac{3t^{1/3}}{r}}}{1 -
c_1e^{\frac{3t^{1/3}}{r}}}\right) dr.\end{equation} Numerical
integration of (\ref{nolene74}) (with $c_1(r)=-1$) shows that the proper radius is a
decreasing function of time. This is expected as the star is
contracting and losing mass. From Figure \ref{fig3} we note that
the proper radius decreases at a faster rate during the latter
stages of collapse.

\section{Discussion}

We have successfully provided an analytical model of a radiating
star undergoing gravitational collapse with non-vanishing shear.
This model has a Friedmann-like limit when the heat flux
vanishes. We further showed that the causal temperature
(representing the stellar fluid out of hydrostatic equilibrium) is
higher than the noncausal temperature at all points of the star.
Further analysis revealed that the relaxation time for the shear
stress (taken to be proportional to the mean collision time)
increases radially outwards, towards the surface of the star. This
is expected, as the outer layers of the fluid are cooler than the
central regions. Of particular significance is the result that the
relaxation time for the heat flux (in our case taken to be
constant) differs from the relaxation time for the shear stress.
This is contrary to earlier treatments where it was assumed that
$(\tau_{r})_{heat}  \approx (\tau_{r})_{shear}$\cite{mart1,herr}.

As presented above, the solution (\ref{line}) together with (\ref{efe1})--(\ref{efe4}) admits singularities at 
$t=0, r=0$ and $1\pm c_1(r)\exp(3t^{1/3}/r)=0$.  The first singularity is avoided by noting that the life of the star is usually taken to start at $t=-\infty$ and end at $t=0$.  The model we present here is really suitable for the early life of a star.  In fact, one can show that a black hole arises as $t\longrightarrow 0$.  This consideration also takes care of the third case.  For early times (large negative values of $t$ in which we only take real roots) $\exp(3t^{1/3}/r)<<1$ and so the expression does not evaluate to zero regardless of the sign taken.  

The singularity at $r=0$ can be avoided if the solution presented is viewed as an ``envelope'' with a core represented by a metric of the form
\begin{equation} ds^2 = - A_0(r)^2 dt^2 +B_0(r)^2(f(t)^2 dr^2 + r^2 g(t)^2 d\Omega^2), \label{core} \end{equation}
where $A_0$ and $B_0$ represent a {\it known} static solution (See Govender {\it et al}\cite{sharma} for a further discussion of this approach.).  It is a simple matter\cite{gg2005} to match (\ref{core}) to (\ref{line}) at some inner boundary $r_0$. Since (\ref{core}) is not singular at $r=0$, this singularity is avoided.

To make a more realistic comparison of the relaxation times, one
requires an analytic solution of the causal temperature equation
for non-constant relaxation times. Future work in this direction
will also require the comparison of the various relaxation times
using the truncated and full transport equations.

\end{document}